\documentstyle[aps,pre]{revtex}

\begin{document}
\begin{center}
{\large \bf A general learning algorithm for solving  
optimization problems and its application to the spin glass
problem} \\
 
\vspace*{0.8cm}
 
{ \bf Kan Chen}\\
{ \sl Department of Computational Science}\\
{ \sl National University of Singapore, Singapore 119260}\\

\end{center}
 
\vspace*{0.8cm}
\begin{abstract}
We propose a general learning algorithm for solving optimization problems,
based on a simple strategy of trial and adaptation.
The algorithm maintains a probability distribution of possible solutions
(configurations), which is updated continuously in the learning process.
As the probability distribution evolves, better and better solutions are 
shown to emerge. The performance of  the algorithm is  illustrated by  the
application to the problem of finding the ground state of the Ising spin 
glass. A simple theoretical understanding of the algorithm is also presented.
\end{abstract}

\vspace*{0.5cm}
 
PACS number(s): 07.05.Mh, 75.10.Nr, 02.60.Pn  

\pagebreak

Many problems in science and engineering can be formulated in terms of
optimization problems.  In these problem we often want to find a set 
of optimum values for a set of variables that minimizes a given function of 
the variables. When the number of variables is large, especially in cases 
where there are a number of local minima, an 
exact solution is usually impossible to obtain; the aim is thus to
find near-optimal solutions. 

A few optimization methods, based on ideas from physics and
biology, have been developed recently which lead to rather good 
\underline{general purpose} algorithms \cite{boun} for solving optimization 
problems. They have been successfully  applied to a wide range of practical 
problems. One of them is a  stochastic algorithm known as optimization by 
simulated annealing (OSA) \cite{kir1}.
The method is based on an analogy with thermodynamics, in particular, on the
way in which the atoms in a liquid find their minimum energy configuration 
of a crystal, when the liquid is annealed or cooled slowly.
The algorithm consists of Monte Carlo dynamics performed at 
a sequence of effective temperatures to simulate the effect of annealing. 
The stochastic dynamics allows access to a larger region of
configuration (solution) space than simple ``quenching'' methods,
and thereby helps in reaching a good solution. However, the Monte Carlo 
search is based on evolution of a \underline{single} starting configuration 
and is thus often confined to a limited region of configuration space and 
is consequently not efficient in searching through configuration space.
Many ideas have been proposed to make OSA efficient. For example, the OSA 
based on the multi-canonical sampling technique has proved quite effective 
in the spin glass problem \cite{berg1}, \cite{celik} and the traveling 
salesman problem \cite{lc}.  Another general purpose algorithm which has been 
used extensively is the Genetic Algorithm (GA) \cite{holl}.
The Genetic Algorithm demonstrates the importance of keeping
many configurations (``species'') in the optimization process.
The algorithm mimics the principles of evolution (using crossover, 
mutation, etc. to update the configurations). GAs have been applied 
to a large range of problems in a wide variety of topics. 
To apply GAs to large-size problems, however, a significant number 
of configurations need to be retained (thus imposing memory requirements) 
and in addition, many independent runs may be needed 
to avoid missing out on good solutions. Thus GAs may not be efficient for 
optimizing large-size problems. 

In this paper, we propose a learning algorithm, which 
is as general as the genetic algorithm, but does not require storing 
many configurations explicitly. What is kept and updated in our 
learning algorithm are probability weights associated with each spin 
that enable one to generate new configurations probabilistically, with 
lower energy configuartions favored; this allows many configurations to be 
kept implicitly. The learning process is constructed in such way that 
the evolution of the probability distribution will lead to better and better 
configurations (solutions). An early version of the algorithm has been 
applied to solve the traveling salesman problem \cite{chen}. We first present
the general algorithm and demonstrate analytically that in its simplest 
version the algorithm leads to progressively more optimal solutions. In the
second part of the paper we present explicit results for the problem of
determining the ground state of the Ising spin glass. 
 
To describe the algorithm, let us consider the general optimization
problem of finding the values of a set of variables or parameters 
$\{\lambda_i, \ i=1,2,...n\}$
such that the function $F(\lambda_1, \lambda_2, ..., \lambda_n)$
is a minimum; $F$ represents a (free) energy or an objective function. 
As is done in GAs, we encode the variables 
$\lambda_i$ using binary digits, and write the function as 
$F(\sigma_1, \sigma_2,...,\sigma_N)
$, where $\{\sigma_k, \ k=1,2,...,N\}$ are binary numbers assuming values 
$0$ and $1$. 
Equivalently, we may use spin variables $s_i$ 
with values $-1$ and $1$. Since we will test our algorithm on the spin glass 
problem, we will employ the spin description
from now on. 

The aim of the optimization 
is to find the spin configuration $\{s_i\}$ that gives rise to the
smallest value of $F$. The learning algorithm for this problem
can be formulated as follows. For each spin $s_i$, a weight $w_i$
is assigned. The probability for choosing $s_i=1$ is defined as
$w_i/(1+w_i)$ and the probability for choosing $s_i = -1$ is
then $1/(1+w_i)$ (Initially $w_i$ is set to be 1, so that the probabilities 
of choosing $s_i =1$ and $s_i = -1$ are equal). The basic ingredients in the 
algorithm are \underline{trial and adaptation}: first 
select a configuration $\{s_i\}$; then modify the weights $\{w_i\}$ to
favor configurations with smaller values of $F$. 

\underline{Selection of Spin Configuration}: A configuration is selected 
by choosing the initial configuration $\{s_i\}$ with the probability 
determined by the weights $\{w_i\}$. In the simplest version of the
algorithm, this configuration is used in the evaluation for updating
the weights. The performance of the algorithm can be greatly improved 
by implementing a local optimization 
on the configuration  (changing $s_i$ from 1 to $-1$ and 
vice versa). The resulting configuration after the local optimization
will then be used for the updating (evaluation) of the weights.
The simplest way to perform the local optimization, which is quite general, 
 consists of individual spin flips to lower the value of $F$. More 
sophisticated algorithms can also be used. But they are likely to be 
problem dependent. 

\underline{Evaluation of the configurations obtained and
modification of weights}: Starting from the second trial, 
the configuration obtained in the current trial is 
compared with the configuration obtained in the previous trial. 
Let $\{s_1, s_2, ...\}$ denote the current configuration and 
$\{s'_1, s'_2,..\}$ denote 
the previous configuration;  let the corresponding values of the function 
to be minimized be $F$ and $F'$ respectively.
The comparison of these two configurations leads 
to the modification of weights described by the equation:
\begin{equation} 
\label{updating}
       w_{i}^{new}=w_{i}^{old}
		 e^{-\alpha (F-F')(s_i -s'_i)/2}, l=1,...,N,
\end{equation}
where N is the total number of spins in the systems, and   
$\alpha$ represents the modification rate of weights (learning rate
of the algorithm). According to this 
rule, if the functional value of the current configuration, $F$, is lower
 (higher) than the previous one, $F'$,  the weight will be modified to favor
(disfavor) the current spin configuration. The new weights will be used in 
the selection of the next configuration according to the
prescription given earlier and the procedure repeated. As learning advances, 
the spin configuration will be gradually ``frozen'' into a near-optimal
configuration. 
 
We study the simplest version of the model
analytically to understand how the learning process leads to better 
solutions. We consider the limit where the
learning rate $\alpha$ is very small, and consequently $w_i$ changes very 
slowly. In the spirit of the ``adiabatic approximation'' 
we can write down an equation describing the change of the weight
as a function of time (defined as the number of trials used):
\begin{equation}
\frac{dw_i}{dt} =\left[\sum_{s_1,...s_N}\sum_{s'_1,
	...,s'_n}P(s_1,...,s_N)P(s'_1,...,s'_n) 
	 \left(  e^{-\alpha(F(s_1,...,s_N)-F(s'_1,...,s'_n))
(s_i-s'_i)/2}-1 \right)\right]w_i,
\end{equation}
where $P(s_1,...,s_N)$ is the probability of generating the 
configuration $\{s_1,...,s_N\}$. Since each spin is chosen independently, 
the probability $P$ can be written as
$$P(s_1,s_2,...,s_N) = \prod_{i}p(s_i),$$
where $p(s_i)$ is the probability that the $i$th
spin in the configuration assumes the value $s_i$.
As we have described earlier, $p(s_i)$ is given by
$w_i/(1+w_i)$ if $s_i=1$ and $1/(1+w_i)$ if $s_i=-1$.
These can be combined into a single expression as follows:
$$p(s_i)=\frac{(1+s_i) w_i/2+(1-s_i)/2}{1+w_i}
	=\frac{1}{2}+\frac{s_i}{2}\frac{w_i-1}{w_i+1}.$$
In the limit $\alpha\rightarrow 0$ we can expand the exponential
function in the above equation and keep only the leading order term.
We have
\begin{equation}
\frac{dw_i}{dt} =  -\alpha w_i\left[\sum_{s_1,...s_N}\sum_{s'_1,
	...,s'_n}P(s_1,...,s_N)P(s'_1,...,s'_n)
	(F(s_1,...,s_N)-F(s'_1,...,s'_n))\frac{(s_i-s'_i)}{2}\right].
\end{equation}
Straightforward manipulations lead to
\begin{equation}
\frac{dw_i}{dt} = -\alpha w_i\sum_{s_1,...s_N}P(s_1,...,s_N)
	F(s_1,...,s_N)(s_i-\frac{w_i-1}{w_i+1}).
\end{equation}
Given the expression for $\frac{dw_i}{dt}$, we can then evaluate
the change in the average functional value as a function
of time. The average is defined as
 $$\bar{F}=\sum_{\{s_i\}}F(\{s_i\})P(s_1,...,s_N),$$
and $ \frac{d\bar{F}}{dt} $ is given by
\begin{equation}
\frac{d\bar{F}}{dt} =
\sum_i\sum_{s_1,...,s_N} F(s_1,...,s_N)P(s_1,...,s_N)\frac{1}{p(s_i)}
	\frac{dp(s_i)}{dt}.
\end{equation}
Now using the fact that
$$\frac{1}{p(s_i)}\frac{dp(s_i)}{dt}=\frac{s_i}{p(s_i)}
	 \frac{1}{(1+w_i)^2}\frac{dw_i}{dt} 
	= \frac{1}{2w_i}(s_i-\frac{w_i-1}{w_i+1})\frac{dw_i}{dt},$$
we obtain
\begin{equation}
\frac{d\bar{F}}{dt}
	=-\frac{\alpha}{2}\sum_i\left[\sum_{s_1,...s_N}P(s_1,...,s_N)
	F(s_1,...,s_N)(s_i-\frac{w_i-1}{w_i+1})\right]^2.
\end{equation}
It can be seen from the above equation that $\bar{F}$ in the
learning process is gradually reduced ($\frac{d\bar{F}}{dt}<0$), and thus 
better and better configurations are obtained as learning advances.
The reduction is diminished when the learning process converges, i.e.,
when the probability of $s_i =1$ approaches one ($w_i$ approaches $\infty$) 
or zero ($w_i$ approaches 0); in these limits  $s_i=\frac{w_i-1}{w_i+1}$
and $\frac{d\bar{F}}{dt}=0$. This result provides some theoretical basis
for our learning algorithm, which is quite valuable as analytical 
understanding of optimization algorithms is typically difficult to obtain. 

We illustrate the performance of the learning
algorithm in the case of the Ising spin glass systems described by the 
Edwards-Anderson Hamiltonian
\begin{equation}
\label{h_def}
	E = - \sum_{<ij>} J_{ij}s_i s_j,
\end{equation}
where the sum includes only the nearest neighbors (4 for two-dimensional
 systems and 6 for three-dimensional systems); the exchange interactions 
$J_{ij}$, between the spins $s_i = \pm 1$, are independent quenched random 
variables which assume the values $\pm 1$ with equal probability. Clearly the
optimization method can be used for other distributions of $J_{ij}$.
The aim of the optimization is to find the spin configuration $\{s_i\}$
with the lowest value of $E$ for a given set of $J_{ij}$, or the total energy 
per site $e$ (defined as $E$ divided by the total number of spins in the 
system). The problem at hand then is to optimize $e(s_1,s_2,...,s_N)$. 

Our studies on the spin glass problems were done using a 200MHz 
SGI Power Challenge and a cluster of SGI Indigo workstations. The
CPU times quoted in this paper have all been converted to the equivalent
CPU times of the SGI Power Challenge. We have studied two-dimensional 
systems of size ranging from $5\times 5$ up to $200 \times 200$ and 
three dimensional systems of size $4 \times 4 \times 4$
up to $25 \times 25 \times 25$. Let us first look at the
performance of the simplest version of the algorithm without local 
optimization. Here we focus on a sample of size $20 \times 20$. The learning
 rate is taken to be $\alpha = 0.1$. Figure 1 shows the energy obtained vs. 
the CPU time $t$ spent (the data are taken every 800 iterations). 
It also shows the lowest energy
obtained up to time $t$. These data clearly show
the convergence of the learning algorithm as described in our 
analysis above. The simple version of the algorithm is certainly 
not efficient as should be expected, but the result obtained is still quite
impressive in view of the simplicity of the algorithm 
(it reduces the energy from around $0$ to about $-1.2$). Better results can 
be obtained using
a smaller learning rate $\alpha$, but this will be more time consuming.

The simplest version of the algorithm can be improved dramatically
even with a simple local optimization where single spin flips are
attempted to lower the energy after the configuration is selected. 
In this local optimization we make a few passes through the
lattice and a spin is flipped if the flipping leads to a lower energy 
configuration; the procedure stops when the configuration can not be 
improved further by flipping any individual spin on the lattice. 
The locally optimized configuration is then used for comparison with the
previous configuration (also locally optimized). It should be mentioned
that this simple local optimization technique does not utilize any
special feature of the spin glass problem; it can thus also be easily 
applied to other optimization problems. Figure 2 shows the lowest
energy obtained vs. the CPU time spent using this technique
(the same $20 \times 20$ sample is used). For comparison we use four 
different learning rates: $\alpha=0.1$,
$\alpha=0.5$, $\alpha=2.5$, and $\alpha=12.5$. It is clear that with 
the use of the single spin flips, the algorithm is made much more efficient.
The dependence of the algorithm on the learning rate is also clearly 
illustrated in the graph. With large $\alpha$, the learning process
converges quickly, but the result obtained is worse than the result
obtained using a slower learning process. There is a tradeoff between 
obtaining a quick solution or a good solution, which is controlled entirely
by the choice of the appropriate value of $\alpha$ in the algorithm. 

In addition to the simple local optimization based on single spin flips,
we can incorporate more sophisticated local optimization techniques in the
algorithm. The use of sophisticated local optimization techniques is likely
to be problem dependent. For the spin glass problem, we use
a local optimization technique similar to the Kernighan-Lin
variable-depth search algorithm used in the graph-partioning problem and
the traveling salesman problem 
\cite{kl}-\cite{papadimitriou}. The idea is to replace the search
for one favorable spin flip by a search for a favorable sequence of
spin flips, using the energy of system to guide the search. A
sequence of spin flips in the variable-depth search 
is obtained \underline{sequentially} as follows. We start with
the spin flip at a selected location in the system and search
its neighbors to find the most favorable spin flip as the
next spin flip in the sequence (the most favorable spin flip
is one that gives rise to the lowest energy among all spin flips considered).
 In general, after the $k$th spin flip the neighbors of all the spins which 
have been flipped will be searched to find the most favorable spin flip 
as the $(k+1)$st spin flip of the
sequence (the flipped spins in the first $k$ spin flips of the sequence 
will not be considered again).

Let $\Delta E(k)$ denote the total 
accumulated change in energy from  the energy of the starting 
configuration due to the $k$ spin flips. To cut short the search that 
most likely will not lead to a better configuration, 
we apply the stopping condition: the search will stop at the $k$th step if 
$\Delta E(k+1)$ is greater than $2d-2$ ($d$ is the spatial dimension). 
We also set a maximum
number of spin flips allowed, denoted by $n$. So the search will stop when 
$k=n$.  After the search
is completed, we choose $k_0$ corresponding to the minimum $\Delta E$.
If $\Delta E(k_0) <0$, then we move the starting configuration to the
energetically better
configuration corresponding to $k=k_0$ (by adopting the first $k_0$
spin flips); otherwise we abandon all the spin flips generated in the search
and keep the original configuration. A new
search will be initiated starting from a new location in the system.

The implementation of the local optimization based on the variable-depth 
search is accomplished by 
keeping a linked list of starting locations for the
search. This is initialized to include all lattice sites at the beginning.
The starting location of the search is taken and removed
from the top of the list. If a better configuration is obtained in the 
search, the locations of all flipped spins together with their 
neighbors will be appended to the list (if they are not already in the 
list). The local optimization will end, if the linked list is empty.

Fig. 3 shows the optimization results based on $n=0$ (no local optimization), 
$n=1$ (single spin flips are used), $n=9$ and $n=100$ (variable-depth
search). The same $20 \times 20$ sample was used. The same learning
rate $\alpha = 2.5$ is used in all the optimizations. It is clear that
better local optimization leads to better overall results for the
learning algorithm. For $n=9$ and $n=100$ the results based on local
optimization alone are already quite close the optimal value (due to
smaller system size), so the improvement due to the learning process 
is small. However for the larger systems the improvement due to the 
learning process can be significant. One advantage of our learning
algorithm is that any local optimization technique can be incorporated easily
into the algorithm. The performance of the local optimization is not
sensitive to the choice of $n$ as long as $n$ is not much 
less than 10. In fact many searches are stopped after the first few steps 
because of the stopping condition we employ. The time taken to perform a
local optimization with $n=100$ is typically $10$ to $15$ times longer
than the time required for a local optimization with only single spin flips. 

The overall performance of the algorithm is summarized in  
Table I (for 2D systems) and Table II (for 3D systems). 
The number of samples we use for each system size ranges from 320 
for the smallest system to 10 for the largest system. The average energy 
(per site) we obtained and the average time taken to reach
the lowest energy configuration are listed, together
with the average number of iterations used. In obtaining the results 
presented in the tables, local optimization with $n=100$
(for 2D systems) and $n=125$ (for 3D systems) was used in
the learning algorithms. To compare with the results obtained in the
literature, we take our best results obtained with $\alpha=0.1$, 
and fit our data using the form $e_L = e_{\infty}+cL^{-d}$
to obtain the energy of the infinite system.
The results are  $e_{\infty}=-1.4028 \pm 0.0019$ for 2D systems 
and $e_{\infty}=-1.7857\pm 0.0026$ for 3D systems. These are consistent with
the best results quoted in the literature \cite{simone}-\cite{berg2}. 
In particular, for two-dimensional systems, Simone et al. \cite{simone} use
an exact algorithm based on the branch and cut technique to find the
ground states of spin glass systems with system size up to $50\times 50$.
They obtain the extrapolated result $e_{\infty}=-1.4022 \pm 0.0003$ using
the same form of the fitting function. It is not clear, however, whether 
their technique can be efficiently implemented for 3D systems or not 
(Finding the ground
state of the 3D spin glass is an NP-complete problem).
For 3D systems, the most efficient algorithm
appears to be the one using a hybrid of
Genetic Algorithm and local optimization. P\`{a}l \cite{pal} 
used the hybrid algorithm
to study 3D systems of sizes up to $14 \times 14 \times 14$ and
he obtained $e_{\infty} = -1.785708\pm 0.000075$ based on the same form of 
the fitting function we used. Our 3D result agrees well with his result.  

Our algorithm is quite fast compared with most 
other algorithms. For example, for $14\times 14 \times 14$ systems, our 
algorithm needs, on average, 1270 seconds with  $\alpha=0.5$ to obtain 
the average energy of $-1.7865$. 
In comparison, the optimization using the hybrid GA implemented by 
P\`{a}l, which itself is much faster than
the original genetic algorithm approach used by Sutton et al.,
 takes, on average, 23540 seconds per run (30 runs were used)
on a 134 MHz SGI Indy computer (which is about four times slower the 
computer we use), to get the same average energy of $-1.7865$. 
If we do not need especially high accuracy, we can choose a larger learning 
rate and obtain the result much more quickly. With $\alpha=12.5$ we can 
obtain the average energy of $-1.7836$ in the average CPU time of $53.6$ 
seconds. We can study systems of size up to $200 \times 200$ and 
$25 \times 25 \times 25$ with $\alpha=12.5$
rather easily without sacrificing much accuracy.

In conclusion, we have demonstrated how a complex optimization 
problem can be solved by a simple learning strategy of trial and
adaptation. The learning process is somewhat similar to the evolution
process in Genetic Algorithms. As in GA this algorithm has the
advantage that it is based on global searches in configuration space. 
 But instead of keeping many configurations explicitly as in GAs we use 
probability weights to generate configurations 
probabilistically. Thus many configurations are implicitly kept for 
effective mutation and crossover through the updating of the 
probability weights. For the simple version of the algorithm without
local optimization, we have shown analytically that the algorithm
does lead to better and better solutions. Local optimization
techniques can also be easily incorporated in the algorithm, which leads to
a rather effective optimization method, as we have demonstrated in the spin
glass problem.  We believe that our learning 
algorithm can be equally effective in other optimization problems,
in particular the ones where sophisticated local search algorithms
have not been found. 

We thank C.\ Jayaprakash for critical reading of the manuscript and
many helpful comments and suggestions. 
 
\pagebreak

\pagebreak
\noindent Figure captions.
 
\vspace{6mm}

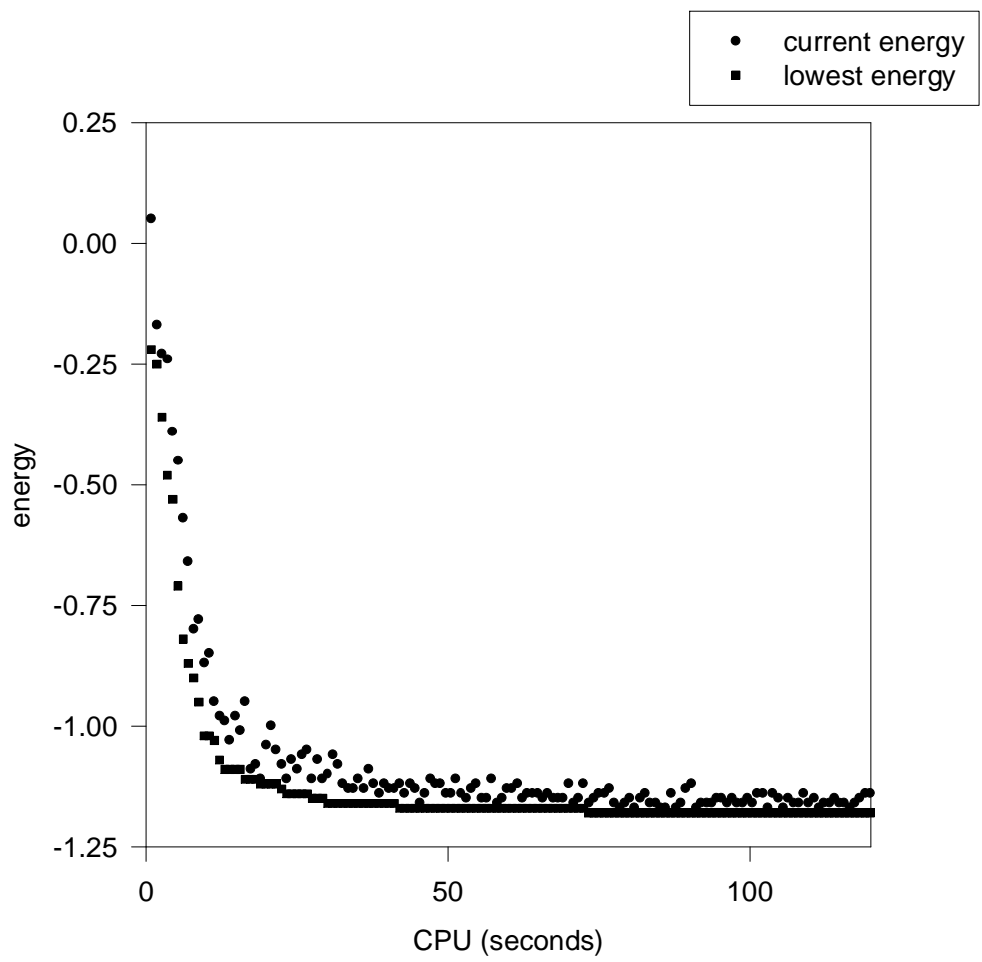
\noindent Figure 1. Energy per site $e$ vs.
the CPU time t used for the optimization of a $20 \times 20$ systems.
The filled-circle represents the energy obtained at
the current time; while the filled square represents the lowest
energy obtained up to time $t$. The data are taken every $800$ iterations,
and the optimization is performed with $\alpha=0.1$.

\vspace{6mm}

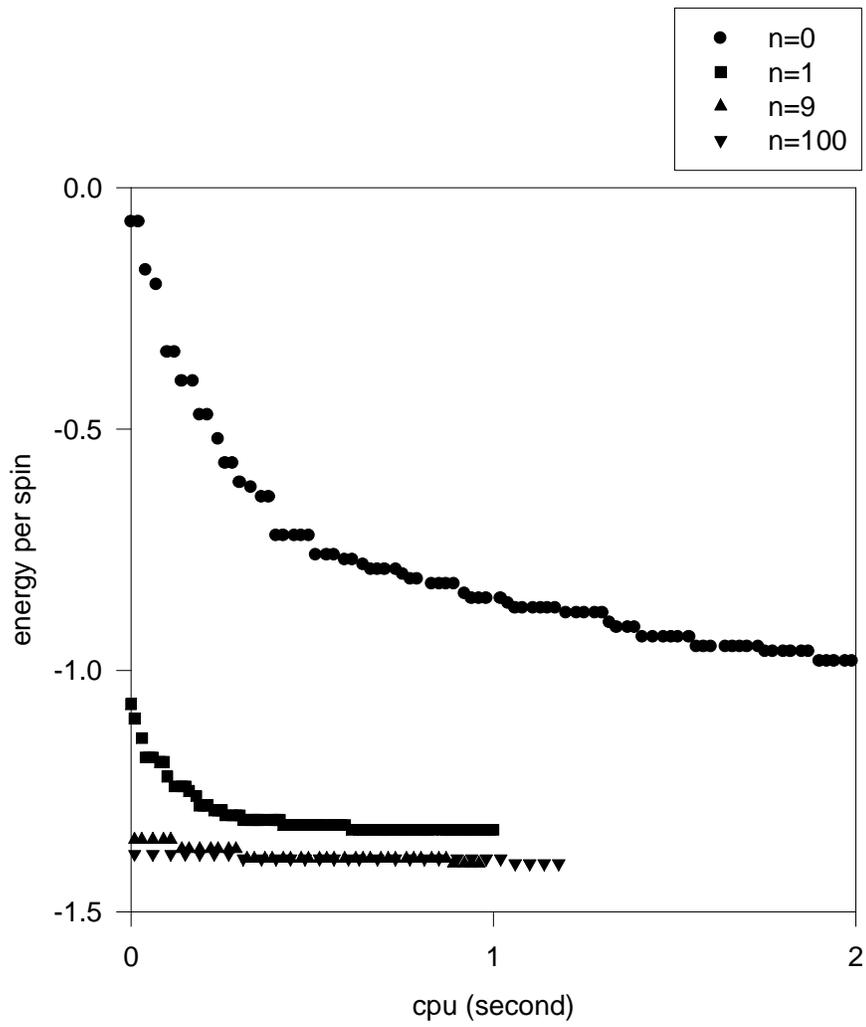
\noindent Figure 2. The lowest energy obtained vs.
the CPU time used, for the optimizations with $\alpha=0.1, 0.5, 2.5$ and 
$12.5$. 
Local optimization with single spin flips are used in the algorithm.
 
\vspace{6mm}

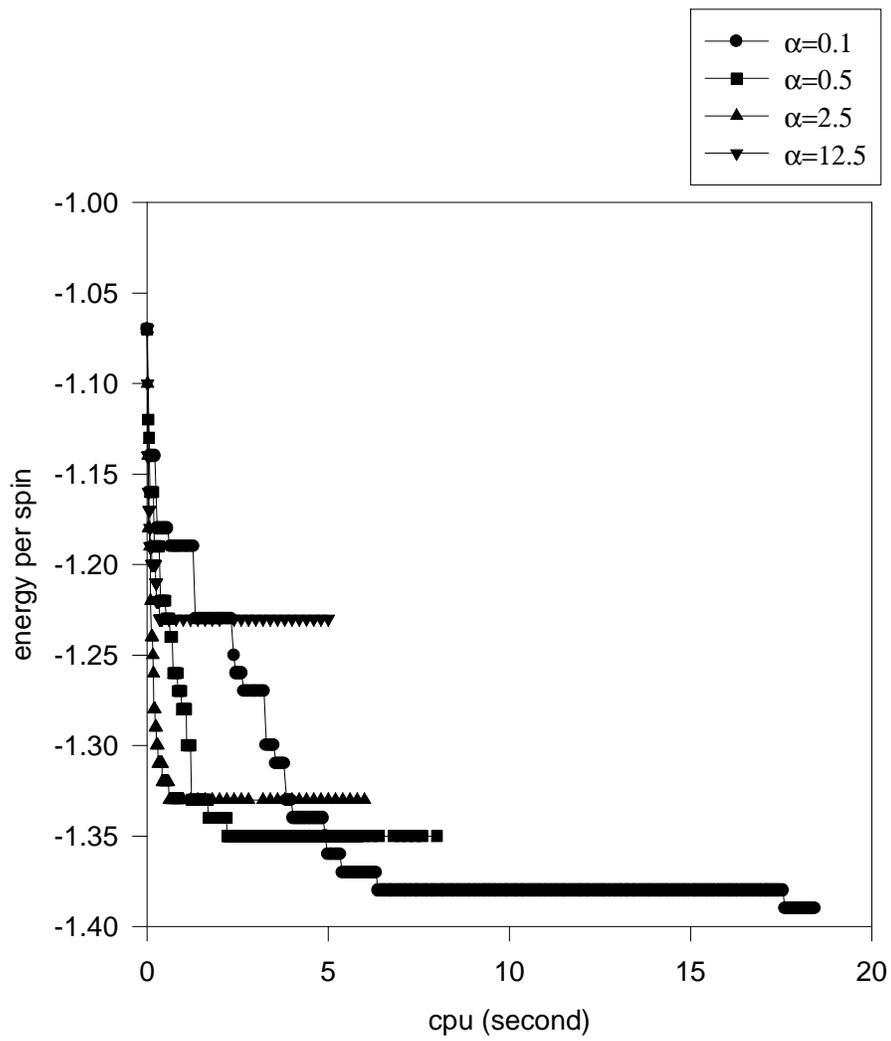
\noindent Figure 3. The lowest energy obtained vs. the CPU time used, for 
the optimizations with $n=0$, $n=1$, $n=9$, and $n=100$. 
 The learning rate for these optimizations is chosen to be $\alpha=2.5$.
The same $20 \times 20$ sample is used.

\vspace{6mm}

\noindent Table captions.
 
\vspace{6mm}
 
\noindent Table I. Average lowest energy, the CPU time,
and the number of trials needed (number in parentheses) 
to reach the lowest energy configuration 
from the optimizations of the two dimensional systems of
size $L \times L$ with $L=5, 10, 20, 30, 40, 50$, and 200.
The values of the learning rate $\alpha$ used are $0.1, 0.5, 2.5$,
and $12.5$. The number of samples $N_s$ used is also listed 
(in parentheses under the system size).

\vspace{6mm}
 
\noindent Table II. Average lowest energy, the CPU time,
and number of trials needed (number in parentheses) 
to reach the lowest energy configuration 
from the optimizations of the three dimensional 
systems of size $L \times L \times L$ with $L=4, 6, 8, 10, 12, 14$, and 25.
The values of the learning rate $\alpha$ used are 0.1, 0.5, 2.5,
and 12.5. The number of samples $N_s$ used is also listed 
(in parentheses under the system size).

\pagebreak
\begin{center}
Table I
\end{center}
 
\begin{tabular}{||c|c|c|c|c||} \hline
\multicolumn{5}{||c||}{Energy per site $e$, convergence time  (seconds),
	and iteration steps} \\ \hline
    L ($N_s$)      &$\alpha=0.1$   & $\alpha=0.5$ & $\alpha=2.5$ 
	& $\alpha=12.5$    \\ \hline
5 &$-1.3405\pm 0.0051$ & $-1.3405\pm 0.0051$ & 
	$-1.3405\pm 0.0051$ & $-1.3405\pm 0.0051$ \\ \cline{2-5}
(320)    & 0.0008 (1.0)& 0.0008 (1.0)&0.0008 (1.0)&0.0008 (1.0) \\ \hline
10 &$-1.3882\pm 0.0037$ & $-1.3882\pm 0.0037$ & 
	$-1.3882\pm 0.0037$ & $-1.3882\pm 0.0037$ \\ \cline{2-5}
(160)    & 0.005 (1.8)& 0.004 (1.7)&0.005 (1.8)&0.004 (1.7) \\ \hline
20 &$-1.4019\pm 0.0022$ & $-1.4019\pm 0.0022$ & 
	$-1.4018\pm 0.0022$ & $-1.4008\pm 0.0022$ \\ \cline{2-5}
(80)    & 0.74 (55)& 0.64 (48)&0.46 (35)&0.23 (20) \\ \hline
30 &$-1.4007\pm 0.0022$ & $-1.4003\pm 0.0022$ & 
	$-1.3999\pm 0.0022$ & $-1.3984\pm 0.0023$ \\ \cline{2-5}
(40)    & 185 (6112)& 51 (1747)&11 (389)&2.2 (82) \\ \hline
40 &$-1.4001\pm 0.0024$ & $-1.4003\pm 0.0026$ & 
	$-1.3995\pm 0.0026$ & $-1.3983\pm 0.0026$ \\ \cline{2-5}
(20)    & 1587 (28670)& 287 (5590)& 49 (955)& 11 (246) \\ \hline
50 &$-1.4002\pm 0.0030$ & $-1.3999\pm 0.0030$ & 
	$-1.3994\pm 0.0030$ & $-1.3988\pm 0.0030$ \\ \cline{2-5}
(10)    & 4503 (53038)& 2959 (9393)&  547(1883)& 150 (579) \\ \hline
200 & &  & 
	& $-1.3976\pm 0.0005$ \\ \cline{2-5}
 (10)   & & &  & 17259(16893) \\ \hline
\end{tabular}

\pagebreak

\begin{center}
Table II
\end{center}
 
\begin{tabular}{||c|c|c|c|c||} \hline
\multicolumn{5}{||c||}{Energy per site $e$, convergence time (seconds),
	and iteration steps} \\ \hline
    L($N_s$)      &$\alpha=0.1$   & $\alpha=0.5$ & $\alpha=2.5$ 
	& $\alpha=12.5$    \\ \hline
4 &$-1.7453\pm 0.0067$ & $-1.7453\pm 0.0067$ & 
	$-1.7453\pm 0.0067$ & $-1.7453\pm 0.0067$ \\ \cline{2-5}
   (320) & 0.006 (1.2)& 0.0006 (1.2)&0.0006 (1.2)&0.0005 (1.2) \\ \hline
6 &$-1.7720\pm 0.0028$ & $-1.7721\pm 0.0027$ & 
	$-1.7721\pm 0.0027$ & $-1.7714\pm 0.0028$ \\ \cline{2-5}
   (160) & 0.094 (5.8)& 0.075 (4.8)&0.077 (5.0)&0.070 (4.7) \\ \hline
8 &$-1.7855\pm 0.0029$ & $-1.7855\pm 0.0029$ & 
	$-1.7849\pm 0.0029$ & $-1.7827\pm 0.0030$ \\ \cline{2-5}
   (80) & 10.8 (208)& 5.1 (109)& 2.2 (50)&0.73 (18) \\ \hline
10 &$-1.7816\pm 0.0021$ & $-1.7811\pm 0.0021$ & 
	$-1.7803\pm 0.0022$ & $-1.7771\pm 0.0021$ \\ \cline{2-5}
   (40) & 263 (3082)& 62.9 (765)& 18.7 (245)& 5.3 (71) \\ \hline
12 &$-1.7816\pm 0.0020$ & $-1.7813\pm 0.0020$ & 
	$-1.7800\pm 0.0024$ & $-1.7780\pm 0.0024$ \\ \cline{2-5}
   (20) & 1253 (9027)& 409 (3208)& 76.9 (609)& 27.9 (236) \\ \hline
14 &$-1.7874\pm 0.0020$ & $-1.7865\pm 0.0020$ & 
	$-1.7871\pm 0.0023$ & $-1.7836\pm 0.0017$ \\ \cline{2-5}
  (10)  & 4305 (20857)& 1270 (6335)& 251 (1312)& 52.6 (261) \\ \hline
25 & &  & 
	& $-1.7815\pm 0.0009$ \\ \cline{2-5}
(10)    & & &  & 10080(11259) \\ \hline
\end{tabular}
 
\end{document}